\documentclass[12pt]{article}
\usepackage{epsfig}
\usepackage{pst-plot,url,epsf}
\usepackage{axodraw}
\usepackage{t1enc}
\newcommand{\beq}{\begin{equation}}
\newcommand{\eeq}{\end{equation}}
\newcommand{\bea}{\begin{eqnarray}}
\newcommand{\eea}{\end{eqnarray}}

\newcommand{\epm}{e^+e^-}
\newcommand{\nn}{\nonumber}

\newcommand{\ra}{\rightarrow}
\def\earr{\end{array}}
\def\barr#1{\begin{array}{#1}}

\begin{document}
\thispagestyle{empty}
\begin{flushright}
Revised version\\
February 2012\\
\vspace*{1.5cm}
\end{flushright}
\begin{center}
{\LARGE\bf Anomalous $Wtb$ coupling and forward-backward asymmetry of 
top quark production at the Tevatron
}\\
\vspace*{2cm}
K. Ko\l odziej\footnote{E-mail: karol.kolodziej@us.edu.pl}\\[1cm]
{\small\it
Institute of Physics, University of Silesia\\ 
ul. Uniwersytecka 4, PL-40007 Katowice, Poland}\\
\vspace*{4.5cm}
{\bf Abstract}\\
\end{center}
An influence of the anomalous $Wtb$ coupling on forward-backward asymmetry 
in top quark pair production at the Tevatron is investigated taking into
account decays of the top quarks to 6 fermion final states containing 
one charged lepton. To this end
the most general effective Lagrangian of the $Wtb$ interaction containing terms
of dimension up to five is implemented into {\tt carlomat}, a general purpose 
Monte Carlo program, which allows to compute automatically all necessary 
cross sections
in the presence of anomalous vector and tensor form factors.
A sample of results which illustrate little effect of the left- and 
right-handed tensor form factors on the $t\bar t$ invariant mass dependent 
forward-backward asymmetry and the charge-signed rapidity distribution of 
the lepton originating from the $W$ boson from top quark decay
is shown.

\vfill

\newpage

\section{Introduction.}

The top quark is the heaviest particle ever observed, with mass close to 
the energy scale of the electroweak symmetry breaking.
Therefore the top quark physics is an ideal place to look for
non-standard effects which may reveal themselves through departures
of the top quark properties and interactions from those predicted by the 
standard model (SM). The observation of a forward-backward 
asymmetry (FBA) in the top quark pair production in high energy 
proton-antiproton collisions at Tevatron \cite{afbCDF}, \cite{afbD0}
that exceeds the SM expectation is an indication that this conjecture 
may be true. The CDF  and D0  Collaborations measured
the total asymmetry $A_{t\bar{t}}$ at a parton-level:
\bea
A_{t\bar{t}}({\rm CDF})=0.158\pm 0.075, \qquad 
A_{t\bar{t}}({\rm D0})=0.196\pm 0.06,\nn
\eea 
which is higher, but not inconsistent with the SM result.
The asymmetry is zero in the lowest order of SM. A small asymmetry 
of $A_{t\bar{t}}=0.06\pm 0.01$ arises at one loop 
QCD in the result of interferences of double-gluon corrections that
differ under charge conjugation \cite{afbQCD}. 
The CDF Collaboration finds that the asymmetry is a rising function of 
the $t\bar{t}$ invariant mass $m_{t\bar{t}}$, with
\bea
A_{t\bar{t}}(m_{t\bar{t}}\geq 450\;{\rm GeV}/c^2)=0.475\pm 0.114, \nn
\eea
which is more than three standard deviations above the SM prediction 
in this $m_{t\bar{t}}$ region \cite{afbCDF}.
The D0 Collaboration measured also a corrected asymmetry based on the lepton
from a top quark decay to be $0.152\pm 0.040$ which should be compared
with the next-to-leading-order Monte Carlo generator result of
$0.021\pm 0.001$ \cite{mcnlo}.
Dedicated analyses of higher order contributions to the FBA of top quarks
in the high 
invariant mass range of $m_{t\bar t} > 450\;{\rm GeV}/c^2$ show that
the inclusion of the higher order QCD \cite{afbNNLO} and electroweak 
\cite{hollik}, \cite{kuehn} corrections increases the one loop QCD prediction 
to some extent, but
a $3\sigma$ deviation between the measurement and the SM prediction 
in this range still remains. 
Several new physics ideas, which alter the SM top quark production mechanism,
have been invoked in order to
explain the discrepancy \cite{afblit}.

At the Tevatron, the top quarks are produced dominantly in pairs through
the quark-antiquark annihilation process
\bea
\label{qqtt}
         q\bar{q} \;\rightarrow\;  t \bar{t}.
\eea
Creation of a top quark pair through the gluon-gluon fusion process,
$gg \rightarrow  t \bar{t}$,
that dominates the top quark production at the LHC, has much smaller 
cross section at the Teavatron. Moreover, it does not contribute 
to the FBA, as its initial state is symmetric under charge conjugation.
Single top production processes, as e.g. $qb\to q't$,
$q\bar{q}'\to t\bar{b}$ or  $qg\to q't\bar{b}$, have much smaller
cross sections at the Tevatron, therefore their possible contribution
to the FBA is neglected in the present work.

\begin{figure}[htb]
\epsfig{file=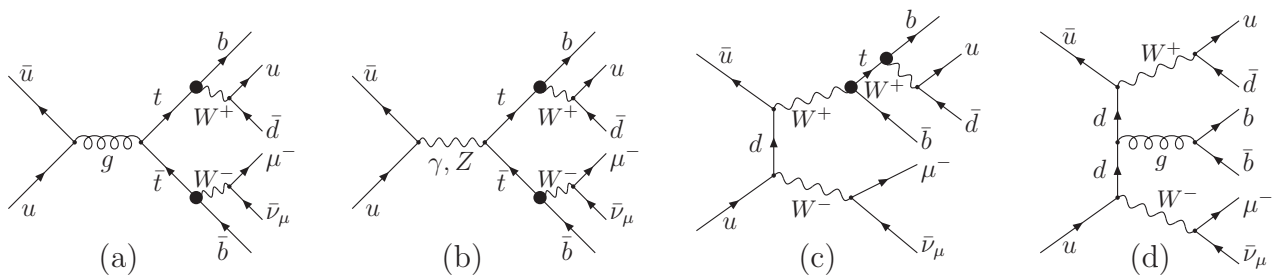,  width=168mm, height=35mm}
\caption{Examples of the lowest order Feynman diagrams of reaction 
(\ref{uu}). Black blobs indicate the $Wtb$ coupling.}
\label{diags}
\end{figure}

Each of the top quarks of reaction (\ref{qqtt})
decays into a $b$ quark and a $W$ boson before hadronization takes place,
and the $W$ bosons decay into a fermion-antifermion pair each.
The top quark pair production at the Teavatron is identified by selecting events
where one $W$ decays to $q\bar{q}'$ and the other to $l\bar{\nu}_l$.
The experimental signature is an isolated electron or muon with large
transverse momentum, a missing transverse momentum from the undetected
neutrino and four or more jets.
At the parton level, one should consider reactions of the form
\bea
\label{tt6f}
u\bar{u} (d\bar{d}) \;\ra\; b q\bar{q}'\;\bar{b} l\bar{\nu}_l,
\eea
where the quark $q$ in the final state may be, but need not be identical 
with the initial state $u$ or $d$ quark. 
Any specific channel of (\ref{tt6f}) 
receives contributions typically from a few hundred Feynman diagrams, 
already at the lowest order of SM.
For example,  in the unitary gauge, assuming vanishing light fermion masses,
$m_u=m_d=m_s=m_e=m_{\mu}=0$, and neglecting the Cabibbo-Kobayashi-Maskawa (CKM)
mixing between quarks, there are 718 lowest order Feynman diagrams for each of 
the reactions
\bea
\label{uu}
u\bar{u} &\ra& b u\bar{d}\; \bar{b} \mu^-\bar{\nu}_{\mu},\\
\label{dd}
\qquad d\bar{d}&\ra&  u\bar{d}\; \bar{b} \mu^-\bar{\nu}_{\mu}.
\eea
Examples of the Feynman diagrams of reaction (\ref{uu}) 
are shown in Fig.~\ref{diags}. They include only six `signal' diagrams
of $t\bar{t}$ production, three of which are depicted in Figs.~\ref{diags}(a) 
and \ref{diags}(b) and the other three are obtained by permutation 
of identical $u$ quarks.
All the remaining diagrams constitute the off resonance background for
the top quark pair production process. Some of them, as the one
shown in Fig.~\ref{diags}(c),  may contain a single 
top quark propagator, but most of the diagrams do not contain the internal
top quark line at all, as the one shown in Fig.~\ref{diags}(d).
Let us note that the $Wtb$ coupling that is indicated by a black blob enters
twice both in the $t\bar t$ production signal
diagrams and the diagram with one top quark propagator. Obviously, it
is not present in the off resonance background diagrams without internal top
quark lines.

The presence of an anomalous $Wtb$ coupling
influences the top quark pair production
in two basic ways. First, it changes the total decay width of the top quark,
which substantially alters the total cross sections of any of reactions
(\ref{tt6f}). Secondly, it changes the differential
distributions of the final state particles, in particular of the final state
lepton, which may have some influence on the $t\bar{t}$ production
event reconstruction.

Therefore, in the present Letter, the anomalous $Wtb$ coupling
of the most general form, with operators up to dimension five, 
is included in the theoretical analysis in order to see to which  extent
its possible modifications 
may change lowest order SM predictions for  the $t\bar t$ invariant 
mass dependent FBA in the top quark pair production at the Tevatron.
The question of whether the anomalous $Wtb$ coupling may affect
the asymmetry based on the charge and rapidity of the
muon originating from the $W$ boson from top quark decay will also be 
addressed.

\section{An anomalous $Wtb$ coupling}

The effective Lagrangian of the $Wtb$ interaction containing operators of 
dimension four and five considered in the present Letter
has the following form \cite{kane}:
\bea
\label{lagr}
L_{Wtb}&=&{g\over\sqrt{2}}\,V_{tb}\left[W^-_{\mu}\bar{b}\,\gamma^{\mu}
\left(f_1^L P_L +f_1^R P_R\right)t 
-\frac{1}{m_W}\partial_{\nu}W^-_{\mu}\bar{b}\,\sigma^{\mu\nu}
  \left(f_2^L P_L +f_2^R P_R\right)t\right]\nn\\
&+&{g\over\sqrt{2}}\,V_{tb}^*\left[W^+_{\mu}\bar{t}\,\gamma^{\mu}
\left(\bar{f}_1^L P_L +\bar{f}_1^R P_R\right)b 
-\frac{1}{m_W}\partial_{\nu}W^+_{\mu}\bar{t}\,\sigma^{\mu\nu}
  \left(\bar{f}_2^L P_L +\bar{f}_2^R P_R\right)b\right],
\eea
where $g$ is the weak coupling constant, $m_W$ is the mass of the $W$ boson, 
$P_{L}=\frac{1}{2}(1-\gamma_5)$ and $P_{R}=\frac{1}{2}(1+\gamma_5)$ 
are the left- and right-handed chirality projectors, 
$\sigma^{\mu\nu}=\frac{i}{2}\left[\gamma^{\mu},
\gamma^{\nu}\right]$, $V_{tb}$ is the element of
the CKM matrix with the superscript * denoting 
complex conjugate, $f_{i}^{L}$, $f_{i}^{R}$, $\bar{f}_{i}^{L}$ and 
$\bar{f}_{i}^{R}$, $i=1,2$, are form factors which can be complex in general. 
There are also other dimension five terms
possible in Lagrangian (\ref{lagr}) for off shell $W$ bosons, 
but they have been neglected as they vanish if the $W$'s decay into massless 
fermions, which is a very good approximation for fermions lighter than 
the $b$ quark. Therefore, in this approximation, Eq.~(\ref{lagr}) represents 
the most general effective Lagrangian of the 
$Wtb$ interaction containing terms of dimension up to five.

The lowest order SM Lagrangian of the $Wtb$ interaction is reproduced 
by setting 
\beq 
\label{sm}
f_{1}^{L}=\bar{f}_{1}^{L}=1, \qquad f_{1}^{R}=f_{2}^{R}=f_{2}^{L}=
\bar{f}_{1}^{R}=\bar{f}_{2}^{R}=\bar{f}_{2}^{L}=0
\eeq
in (\ref{lagr}). 
If $CP$ is conserved 
then the following relationships between
the form factors hold 
\beq
\label{rel}
\left.\bar{f}_1^{R}\right.^*=f_1^R, 
\quad \left.\bar{f}_1^{L}\right.^*=f_1^L, \qquad \qquad
\left.\bar{f}_2^R\right.^*=f_2^L, \quad \left.\bar{f}_2^L\right.^*=f_2^R
\eeq
and 4 independent form factors are left in Lagrangian (\ref{lagr}).
The Feynman rules resulting from Lagrangian (\ref{lagr}) 
are as follows \cite{eewtb}:
\bea
\label{Wtb1}
\barr{l}
\begin{picture}(90,80)(-50,-36)
\Text(-45,5)[lb]{$t$}
\Text(35,27)[rb]{$b$}
\Text(35,-27)[rt]{$W_{\mu}^+, q$}
\Vertex(0,0){2}
\ArrowLine(0,0)(35,25)
\ArrowLine(-45,0)(0,0)
\Photon(0,0)(35,-25){2}{3.5}
\end{picture} 
\earr
\barr{l}
\\[6mm]
\longrightarrow \quad
\Gamma^{\mu}_{t\ra bW^+}={g\over\sqrt{2}}V_{tb}
\left[\gamma^{\mu}\left(f_1^L P_L +f_1^R P_R\right) \right.
                                                                   \\[3mm]
\quad \qquad \qquad \qquad \qquad \qquad \left.-i\frac{q_{\nu}}{m_W}
\sigma^{\mu\nu}\left(f_2^L P_L +f_2^R P_R\right)\right],
\earr
\eea
and
\bea
\label{Wtb2}
\barr{l}
\begin{picture}(90,80)(-50,-36)
\Text(-45,5)[lb]{$\bar{t}$}
\Text(35,27)[rb]{$\bar{b}$}
\Text(35,-27)[rt]{$W_{\mu}^-, q$}
\Vertex(0,0){2}
\ArrowLine(35,25)(0,0)
\ArrowLine(0,0)(-45,0)
\Photon(0,0)(35,-25){2}{3.5}
\end{picture} 
\earr
\barr{l}
\\[6mm]
\longrightarrow \quad
\Gamma^{\mu}_{\bar{t}\ra \bar{b}W^-}={g\over\sqrt{2}}V_{tb}^*
\left[\gamma^{\mu}\left(\bar{f}_1^L P_L + \bar{f}_1^R P_R\right) \right.
                                                                   \\[3mm]
\quad \qquad \qquad \qquad \qquad \qquad \left.-i\frac{q_{\nu}}{m_W}
\sigma^{\mu\nu}\left(\bar{f}_2^L P_L+\bar{f}_2^R P_R\right)\right],
\earr
\eea
where $q$ is a four momentum of the $W$ boson outgoing from the $Wtb$ vertex.

Direct Tevatron limits, that have been
obtained by investigating two form factors at a time
and assuming the other two at their SM values, are the
following \cite{wtbCDF}\footnote{After this work had been submitted
for publication new one-dimensional direct constraints at 95\% C.L. 
on the form factors were  announced by the D0 Collaboration \cite{wtbD0}:
$\left|V_{tb}f_{1}^R\right|^2 < 0.93$, 
$\left|V_{tb}f_{2}^R\right|^2 < 0.13$, 
$\left|V_{tb}f_{2}^L\right|^2 < 0.06$.}:
\begin{eqnarray}
\label{limits}
\left|f_{1}^R\right|^2 < 1.01, \qquad
\left|f_{2}^R\right|^2 < 0.23, \qquad
\left|f_{2}^L\right|^2 < 0.28.
\end{eqnarray}
The direct LHC limits that have been discussed 
in \cite{wtbLHC} are still weaker. 
If $CP$ is conserved then the right-handed vector coupling and tensor couplings 
can be indirectly constrained from the CLEO data on
$b\rightarrow s\gamma$ \cite{cleo} and from other rare $B$ 
decays \cite{fajfer}. However, there is 
still some room left within which the anomalous form factors
can be varied, in particular the tensor ones.

The anomalous $Wtb$ couplings (\ref{Wtb1}) and (\ref{Wtb2}) are implemented 
into {\tt carlomat}, a general purpose program for Monte 
Carlo (MC) computation of lowest order cross sections \cite{carlomat}. 
A new version of the program obtained in this way
allows to make predictions for the top 
quark production and decay through different possible
partonic subprocesses while taking into account complete sets of the lowest
order Feynman diagrams and full information on spin
correlations between the top quark and its decay products. 
The new version of {\tt carlomat} can also
be applied for studying anomalous effects in the top quark production and decay
at the LHC, or 
in $\epm$ collisions at a linear collider \cite{ILC}, \cite{CLIC}. 

\section{Results}

In this section, a sample of results that illustrate the influence
of the tensor form factors of anomalous $Wtb$ couplings (\ref{Wtb1})
and (\ref{Wtb2}) on the $t\bar{t}$ invariant mass dependent 
asymmetry in the top quark production and on
the charge-signed muon rapidity distribution at the high energy
$p\bar{p}$ collisions at the Tevatron is shown. 
The results have been obtained with the current version of {\tt carlomat}.

The physical input parameters that are used in the computation are 
the following: the gauge boson masses and widths
\bea
m_W=80.419\; {\rm GeV}, \quad \Gamma_W=2.12\; {\rm GeV}, \qquad
m_Z=91.1882\; {\rm GeV}, \quad \Gamma_Z=2.4952\; {\rm GeV},
\eea
the heavy quark masses and the Higgs boson mass
\bea
m_t=172.5\; {\rm GeV},  \qquad m_b=4.4\; {\rm GeV}, \qquad
 m_H=115\;{\rm GeV}
\eea
and the coupling constants
\bea
\alpha_W = 1/132.5049458,\qquad \alpha_s(m_Z) = 0.118.
\eea
The QCD couplings are parametrized by $g_s=\sqrt{4\pi\alpha_s}$. The
electroweak coupling constants are parametrized in terms of
$g=\sqrt{4\pi\alpha_W}$
and the complex electroweak mixing parameter
$\sin^2\theta_W=1-{M_W^2}/{M_Z^2}$,
with the complex masses of the $W$ and $Z$ bosons
$M_V^2=m_V^2-im_V\Gamma_V$, $V=W, Z$. The complex gauge boson masses
together with the complex masses of the Higgs boson and top quark 
$M_H^2=m_H^2-im_H\Gamma_H$ and $M_t=\sqrt{m_t^2-im_t\Gamma_t}$, 
where ${\rm Re}\,M_t>0$,
replace masses in the corresponding propagators, both in the $s$- 
and $t$-channel Feynman diagrams.
This choice of parametrizations  is referred to as the
{\em `complex mass scheme'} in {\tt carlomat}.
The Higgs boson width is fixed at the lowest order SM value 
$\Gamma_H=4.9657$~MeV
and the width of the top quark is calculated to the lowest order
with effective Lagrangian (\ref{lagr}) for any specific choice of 
the form factors.

The $t\bar{t}$ invariant mass dependent forward-backward asymmetry 
$A_{t\bar{t}}$ is defined by
\bea
\label{afb}
A_{t\bar{t}}(m_{t\bar{t},i})=\frac{\sigma(\Delta y > 0,m_{t\bar{t},i})
-\sigma(\Delta y > 0,m_{t\bar{t},i})}
{\sigma(\Delta y > 0,m_{t\bar{t},i})+\sigma(\Delta y > 0,m_{t\bar{t},i})},
\eea
with $\Delta y =y_t-y_{\bar{t}}$ being a difference of rapidities
of the $t$ and $\bar t$ quarks with their invariant mass $m_{t\bar t}$
within $i$-th bin.
Since $\Delta y$ is independent of boosts along the beam axis, asymmetry
(\ref{afb}) can be regarded as measured in the $t\bar{t}$ centre of mass system.

The cross section of top quark pair production in $p\bar{p}$ collisions 
at $\sqrt{s}=1.96$~TeV of Eq.~(\ref{afb}) 
is calculated by folding CTEQ6L parton distribution 
functions \cite{CTEQ} with the cross section of hard scattering subprocess
of the top quark pair production of the form (\ref{tt6f}), 
including all the subprocesses
with $u\bar u$ and $d\bar d$ in the initial
state and a single charged lepton in the final state, as e.g. processes
(\ref{uu}) and (\ref{dd}).
The factorization scale is assumed to be equal to a square of the reduced 
centre of mass system energy, $\hat{s}=x_1x_2s$, with $x_1$ ($x_2$)
being a fraction of energy carried by the initial state quark (antiquark).
The $t\bar{t}$ production events are identified with the following 
acceptance cuts on the transverse momenta $p_T$, pseudoprapidities $\eta$, 
missing transverse energy $/\!\!\!\!E^T$ and separation 
$\Delta R_{ik}=\sqrt{\left(\eta_i-\eta_k\right)^2
+\left(\varphi_i-\varphi_k\right)^2}$ in the 
pseudorapidity--azimuthal angle $(\varphi)$ plane between 
the objects $i$ and $k$:
$$p_{Tl} > 50\;{\rm GeV}/c, \qquad p_{Tj} > 50\;{\rm GeV}/c, \qquad
\left|\eta_l\right| < 2.0, \qquad \left|\eta_j\right| < 2.5,$$
\bea
\label{cuts}
/\!\!\!\!E^T > 20\;{\rm GeV},\qquad
\Delta R_{ll,lj,jj} > 0.4.
\eea
The subscripts $l$ and $j$ in (\ref{cuts}) stand for {\em lepton} 
and {\em jet}, a direction of the latter being identified with the 
direction of the corresponding quark. Cuts (\ref{cuts}) are rather restrictive.
It has been checked by a direct computation that 
only events with the invariant masses of  $b q\bar{q}'$ and 
$\bar{b} l\bar{\nu}_l$ subsystems each close to $m_t$ survive. 
This means that the off resonance background contributions are heavily
suppressed and asymmetry (\ref{afb}) is in practice dominated by 
the events of $t\bar{t}$ production and decay.

For the sake of simplicity, it is assumed
that $V_{tb}$ and form factors $f_{i}^{L}$, $f_{i}^{R}$, $\bar{f}_{i}^{L}$ and 
$\bar{f}_{i}^{R}$, $i=1,2$ of Lagrangian (\ref{lagr}) are real. 
As the global fit combined with the SM constraints gives 
$\left|V_{tb}\right|=0.999152^{+0.000030}_{-0.000045}$ \cite{PDG}, 
a value of $V_{tb}$ is fixed at $V_{tb}=1$. Moreover, the vector form factors
are assumed at their SM values of (\ref{sm}),
$f_{1}^{L}=\bar{f}_{1}^{L}=1$, $f_{1}^{R}=\bar{f}_{1}^{R}=0$
and only the tensor form factors are being varied. 

\begin{figure}[!tb]
\vspace{100pt}
\begin{center}
\setlength{\unitlength}{1mm}
\begin{picture}(35,35)(0,0)
\includegraphics{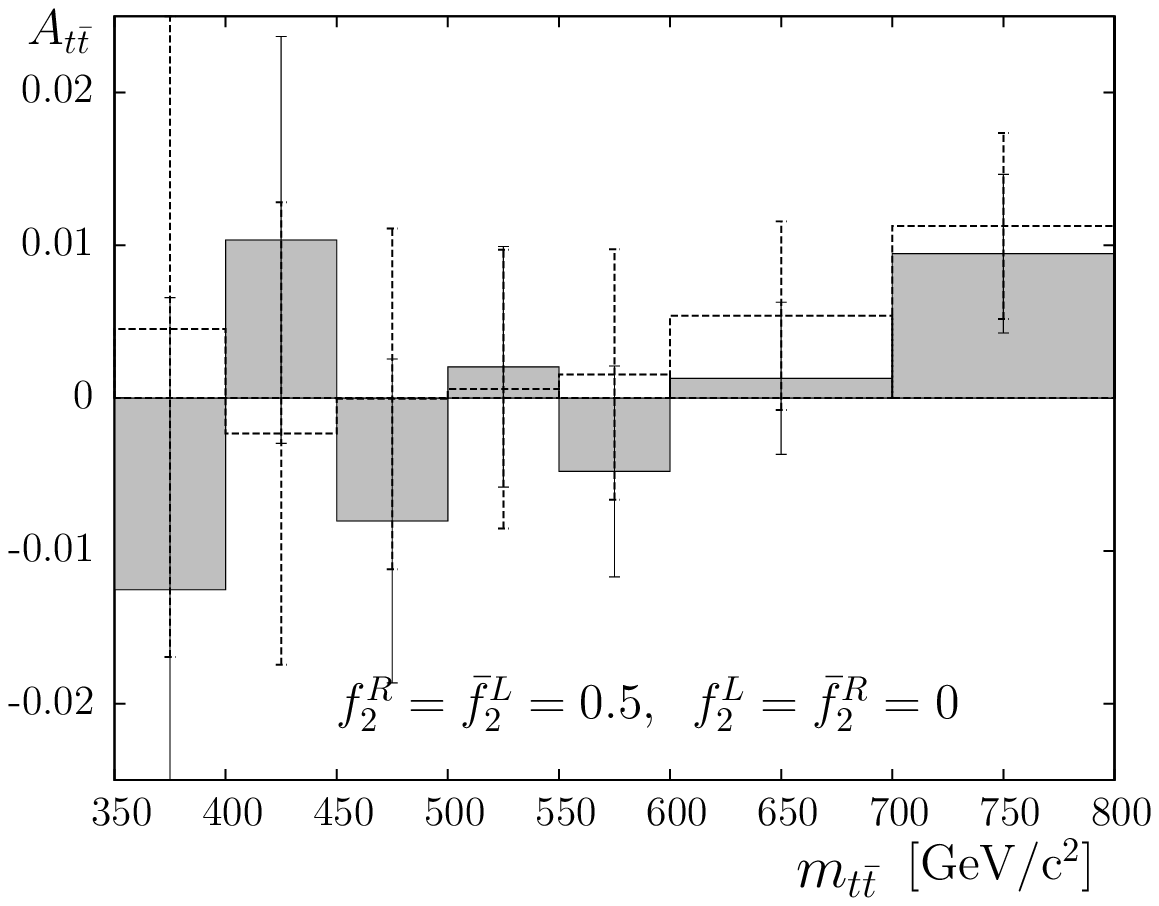}
\end{picture}
\hfill
\begin{picture}(35,35)(0,0)
\includegraphics{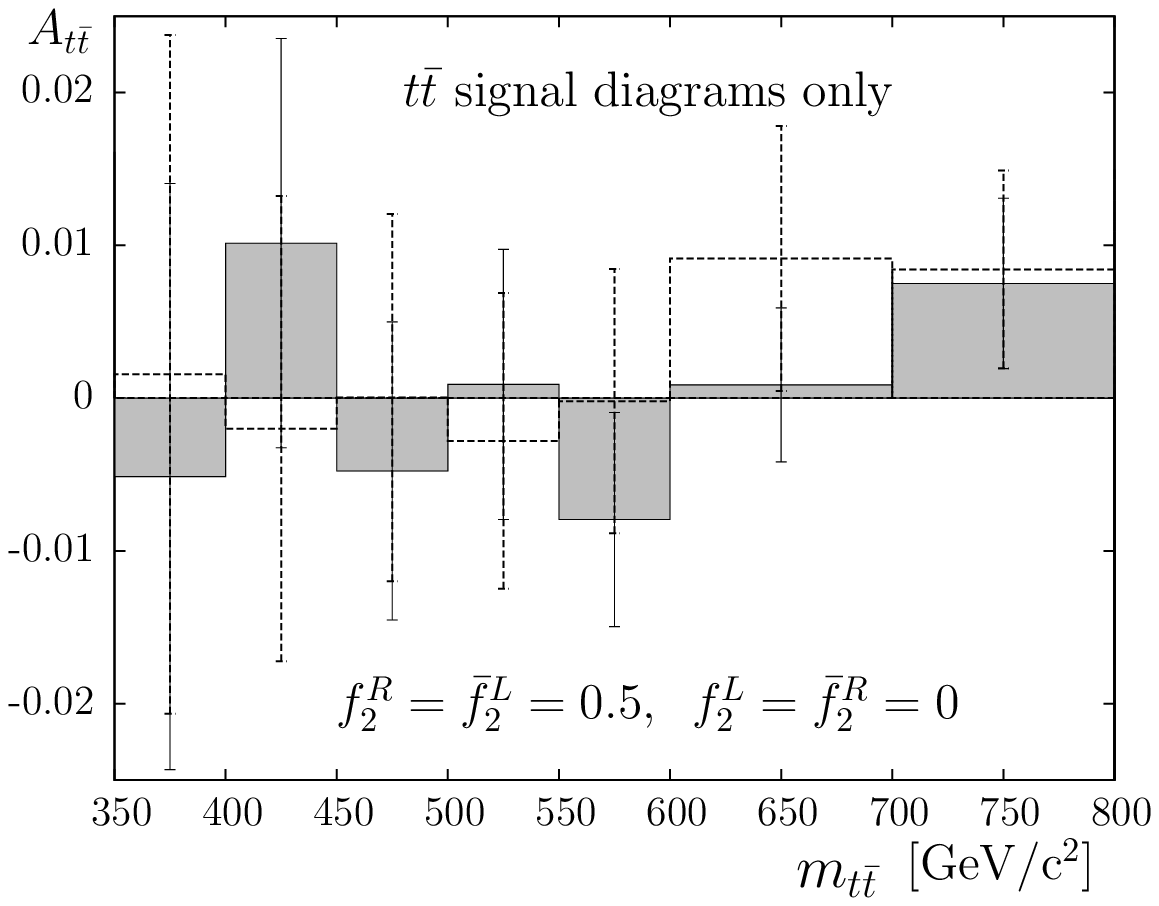}
\end{picture}\\[2cm]
\begin{picture}(35,35)(0,0)
\includegraphics{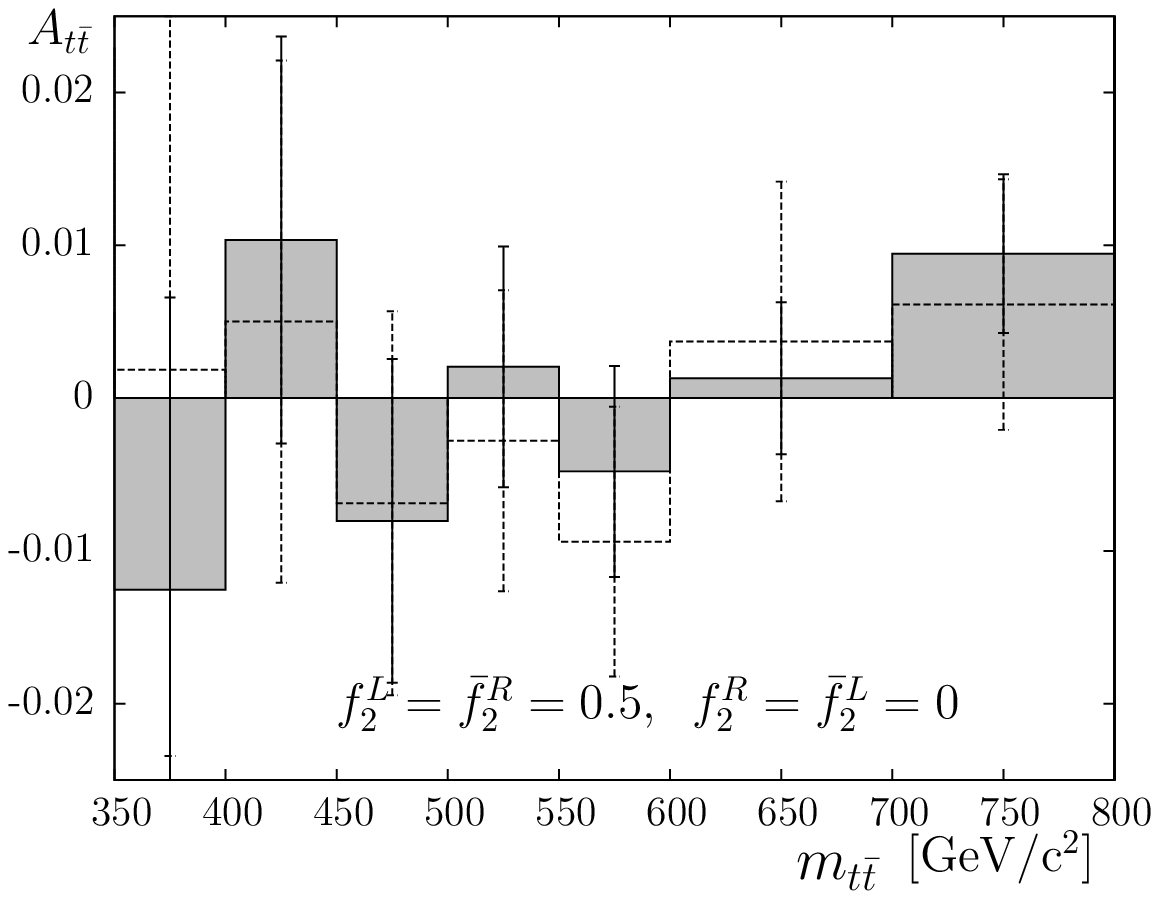}
\end{picture}
\hfill
\begin{picture}(35,35)(0,0)
\includegraphics{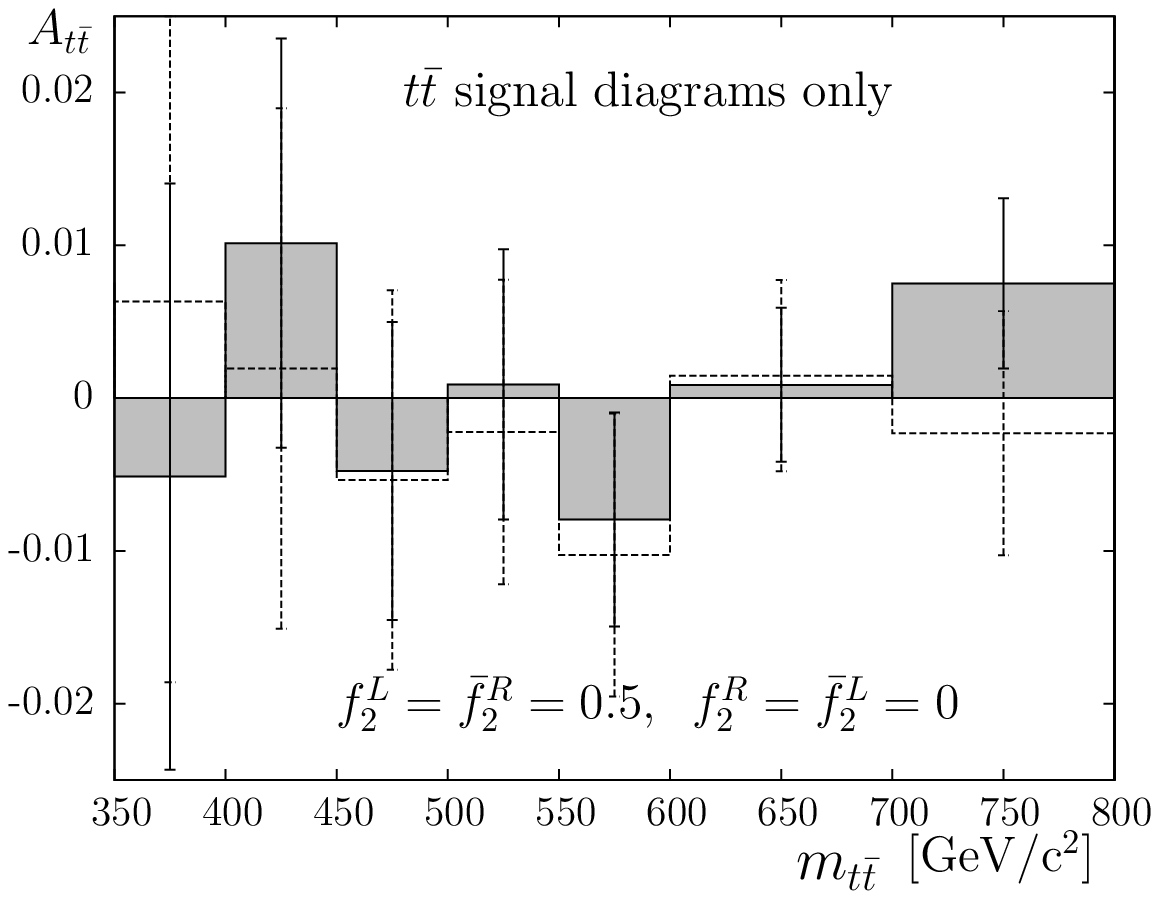}
\end{picture}
\end{center}
\vspace*{-1.5cm}
\caption{Asymmetry (\ref{afb}) calculated with the complete set of the lowest 
order Feynman diagrams (left) and the $t\bar t$ signal diagrams only (right)
for two different $CP$-even choices of the tensor form factors of
(\ref{lagr}). The vector form factors are fixed at their SM values of 
Eq.~(\ref{sm}). The asymmetry for all the form factors fixed by Eq.~(\ref{sm})
is shown in grey in each panel.
}
\label{figafb}
\end{figure}

In Fig.~\ref{figafb}, asymmetry (\ref{afb}) is plotted 
as a function of $m_{t\bar{t}}$, in bins of 50 GeV/$c^2$ below 600 GeV/$c^2$ 
and 100 GeV/$c^2$ above that.
The plots in panels on the left hand side have been obtained
with the complete set of the lowest order Feynman diagrams of each
contributing subprocess and
those in panels on the right hand side have been obtained
with the $t\bar t$ signal Feynman diagrams only.
The result that corresponds to
the form factors satisfying lowest order SM relations (\ref{sm}) 
is depicted with grey boxes in each panel, with solid error bars showing  
one standard deviation of the MC integration in separate bins.
Boxes bounded by dashed lines show the asymmetry in the presence of
two $CP$-even combinations of tensor form factors:
$f_2^R=\bar{f}_2^L=0.5$, $f_2^L=\bar{f}_2^R=0$ in the upper raw panels
and $f_2^L=\bar{f}_2^R=0.5$, $f_2^R=\bar{f}_2^L=0$ in the lower raw panels.
The error bars drawn with dashed lines show the corresponding one 
standard deviation of the MC integration in separate bins.
There are some fluctuations visible in separate bins, but they do not
exceed $2\sigma$. Also the 
total lowest order asymmetry computed with {\tt carlomat}
is consistent with zero within one standard deviation of the MC
integration, as it is shown in Table \ref{Tab:afb}. The asymmetry plots for
other combinations of the tensor form factors, including the $CP$-odd ones, 
look very similar, so they are not shown here. Needless to say, the effect
of the anomalous form factors becomes smaller if they are
chosen within the recent D0 limits \cite{wtbD0}.

\begin{table}[!ht]
\begin{center}
\begin{tabular}{cccc}
\hline 
\hline 
Form factors &
\multicolumn{1}{c}{\rule{0mm}{7mm} 
$m_{t\bar t}< 450\;{\rm GeV}/c^2$} &
\multicolumn{1}{c}{$m_{t\bar t}\geq 450\;{\rm GeV}/c^2$} &
\multicolumn{1}{c}{Total}\\ [1.5mm]
\hline 
\multicolumn{1}{c}{\rule{0mm}{7mm}
$f_2^R=f_2^L=\bar{f}_2^R=\bar{f}_2^L=0$}&
$0.09\pm 1.11 $ & $-0.27\pm 0.45$ & $-0.13\pm 0.52$ \\[1.5mm]
\hline 
\multicolumn{1}{c}{\rule{0mm}{7mm}
$f_2^R=\bar{f}_2^L=0.5,\;f_2^L=\bar{f}_2^R=0$}&
$0.07\pm 1.23 $ & $0.17\pm 0.50$ & $ 0.13\pm 0.54$\\[1.5mm]
\hline 
\multicolumn{1}{c}{\rule{0mm}{7mm}
$f_2^L=\bar{f}_2^R=0.5,\;f_2^R=\bar{f}_2^L=0$}&
$0.39\pm 1.45$ & $-0.42\pm 0.58$ & $-0.09\pm 0.68$\\[1.5mm]
\hline
\hline
\end{tabular}
\end{center}
\caption{Asymmetry (\ref{afb}) in \% integrated in 
different $t\bar t$ invariant mass ranges. The vector form factors are 
fixed at their SM values of Eq.~(\ref{sm}). The complete set of the lowest 
order Feynman diagrams is included.}
\label{Tab:afb}
\end{table}

\begin{figure}[!tb]
\vspace*{-1cm}
\begin{center}
\setlength{\unitlength}{1mm}
\begin{picture}(35,35)(0,0)
\includegraphics{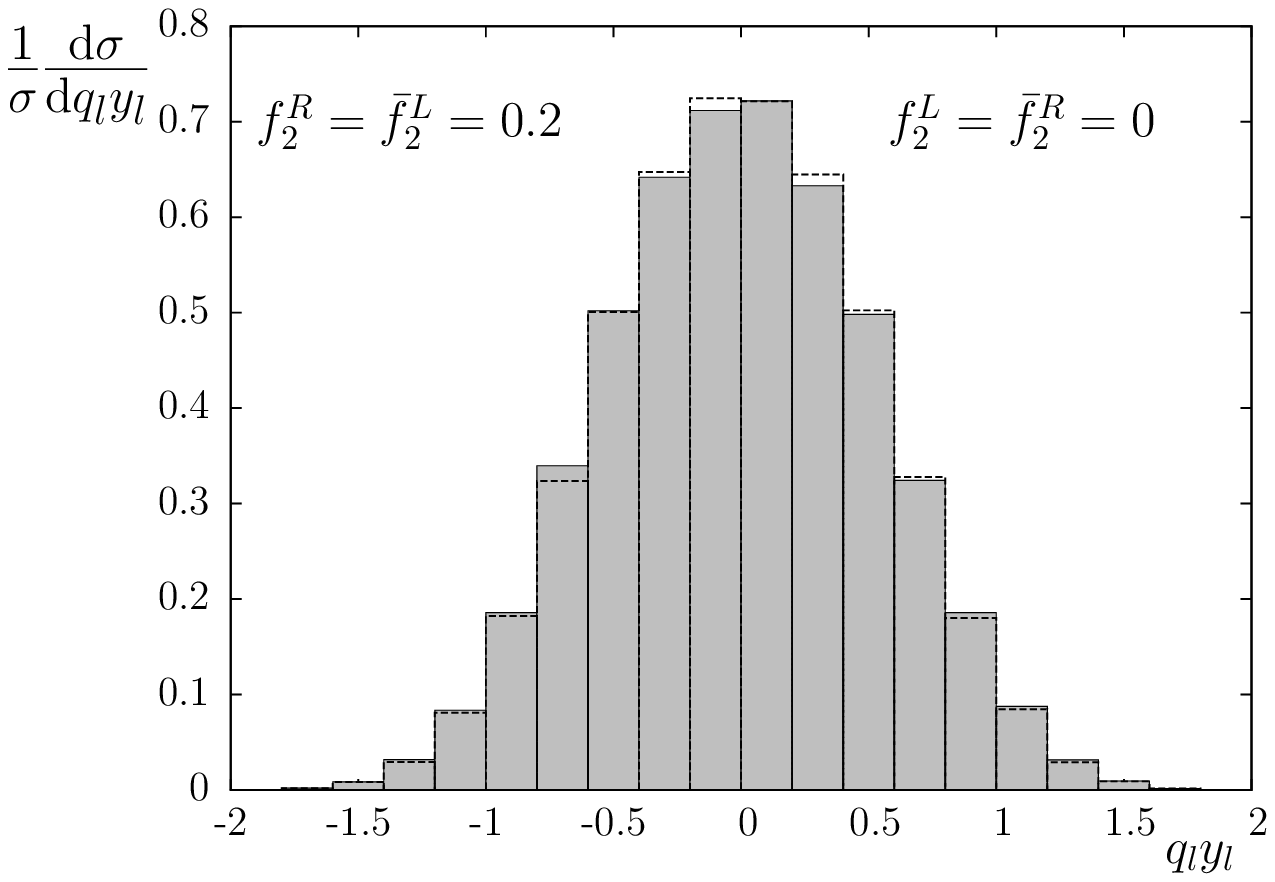}
\end{picture}
\hspace*{4.cm}
\begin{picture}(35,35)(0,0)
\includegraphics{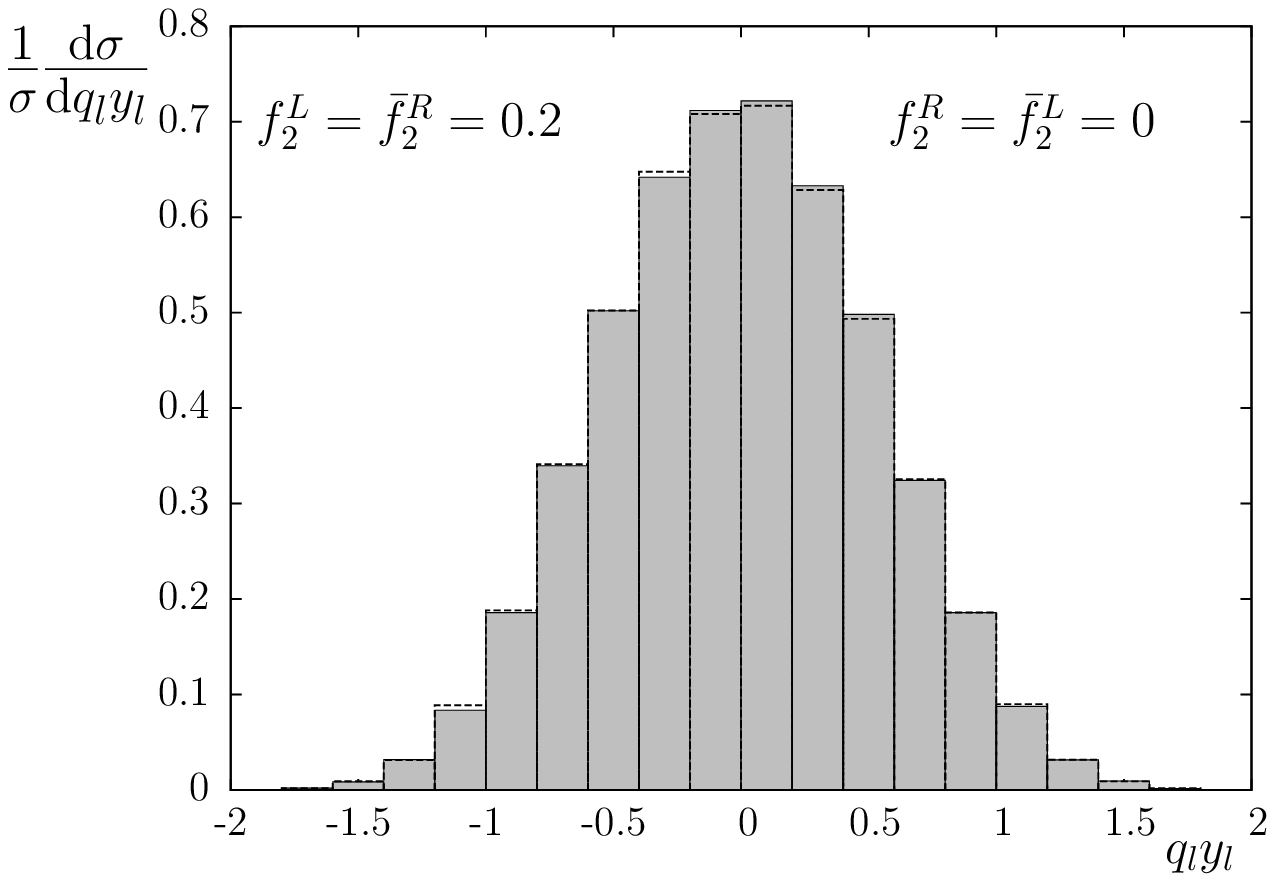}
\end{picture}
\hfill
\begin{picture}(35,35)(0,0)
\includegraphics{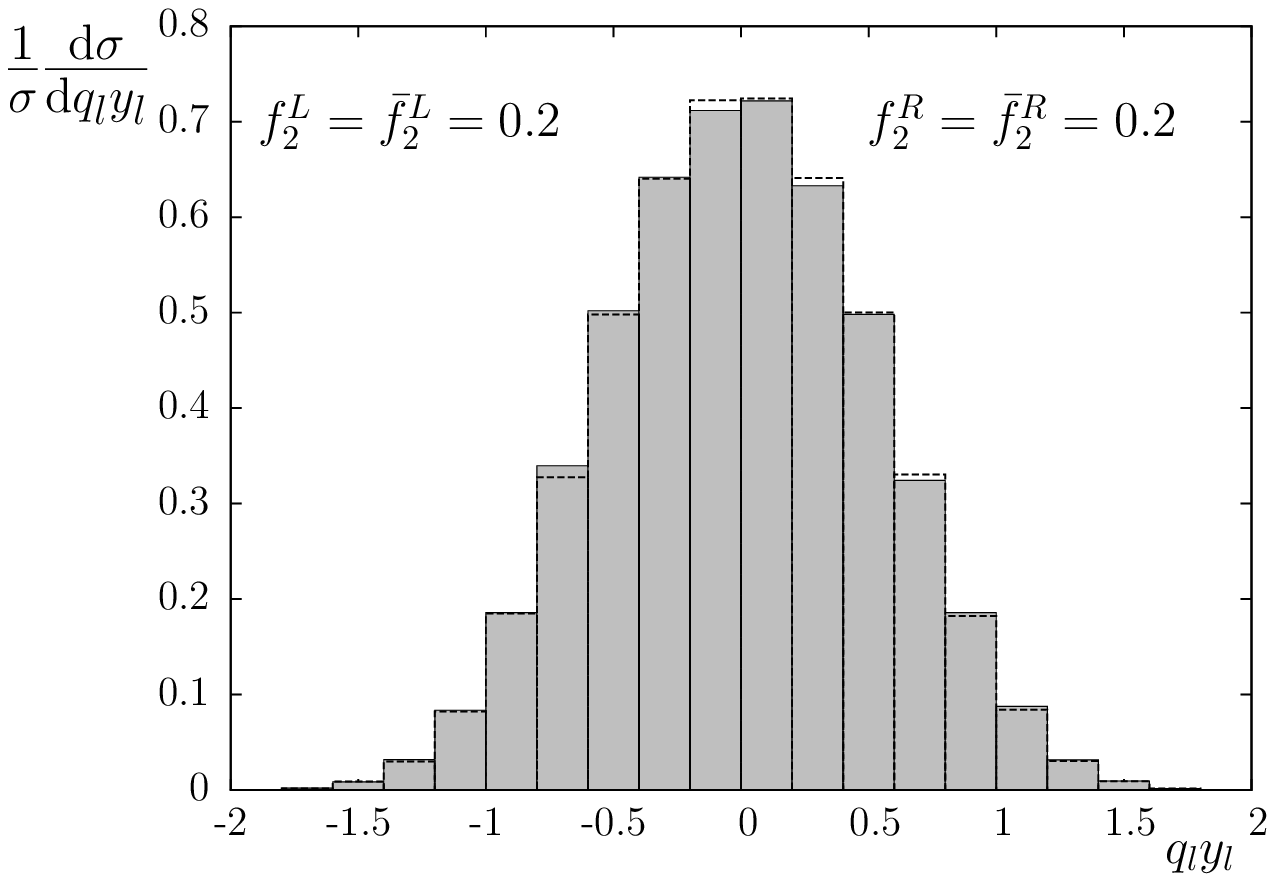}
\end{picture}\\[0.2cm]
\begin{picture}(35,35)(0,0)
\includegraphics{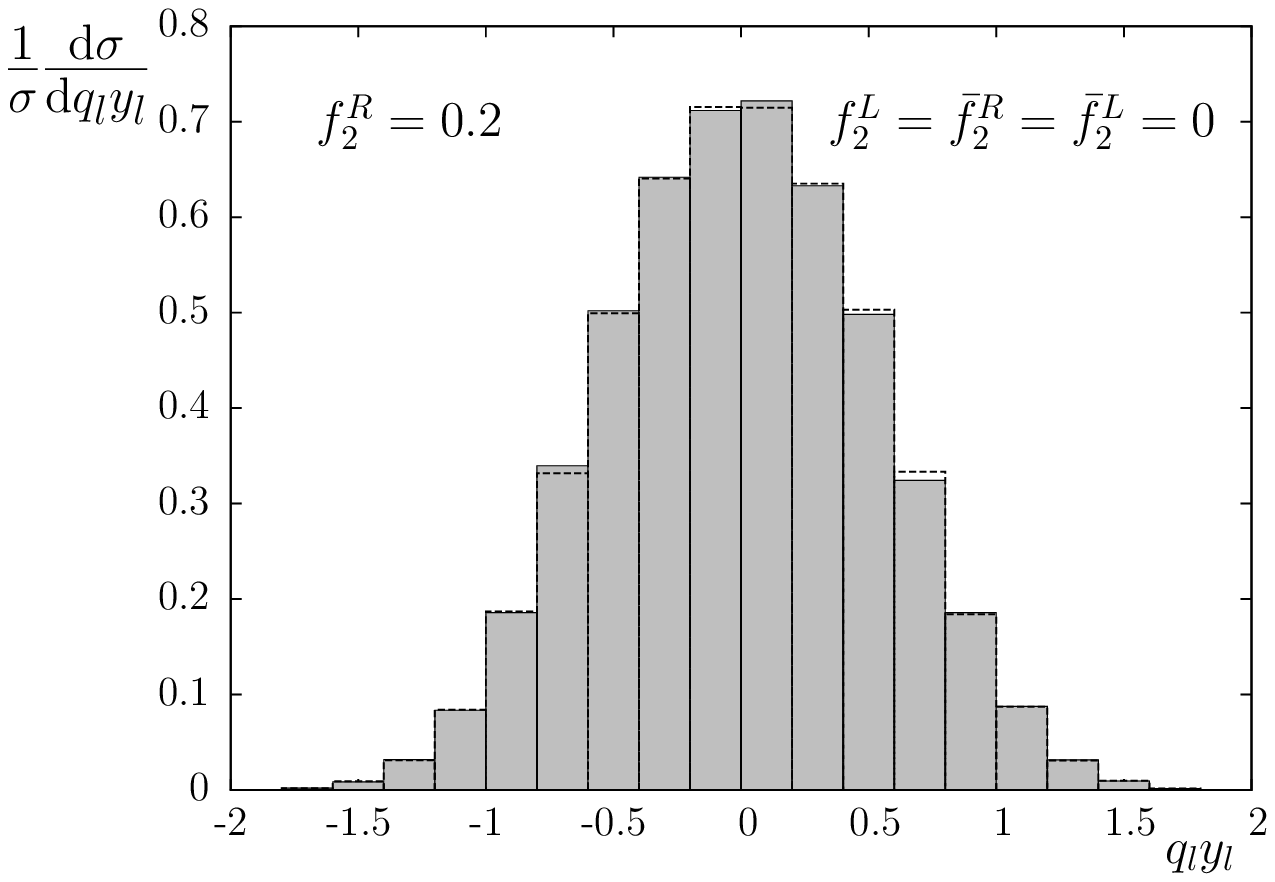}
\end{picture}
\hspace*{4.cm}
\begin{picture}(35,35)(0,0)
\includegraphics{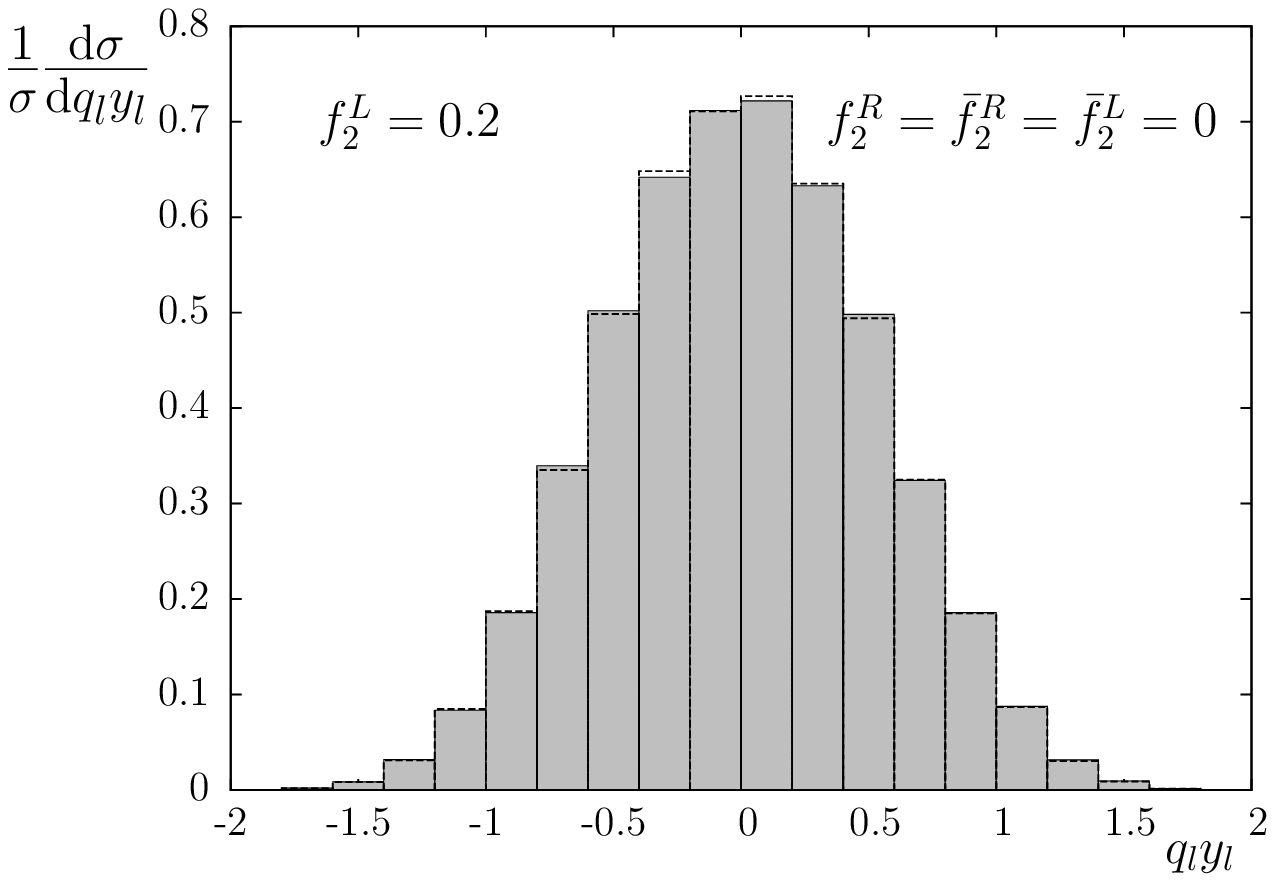}
\end{picture}
\hfill
\begin{picture}(35,35)(0,0)
\includegraphics{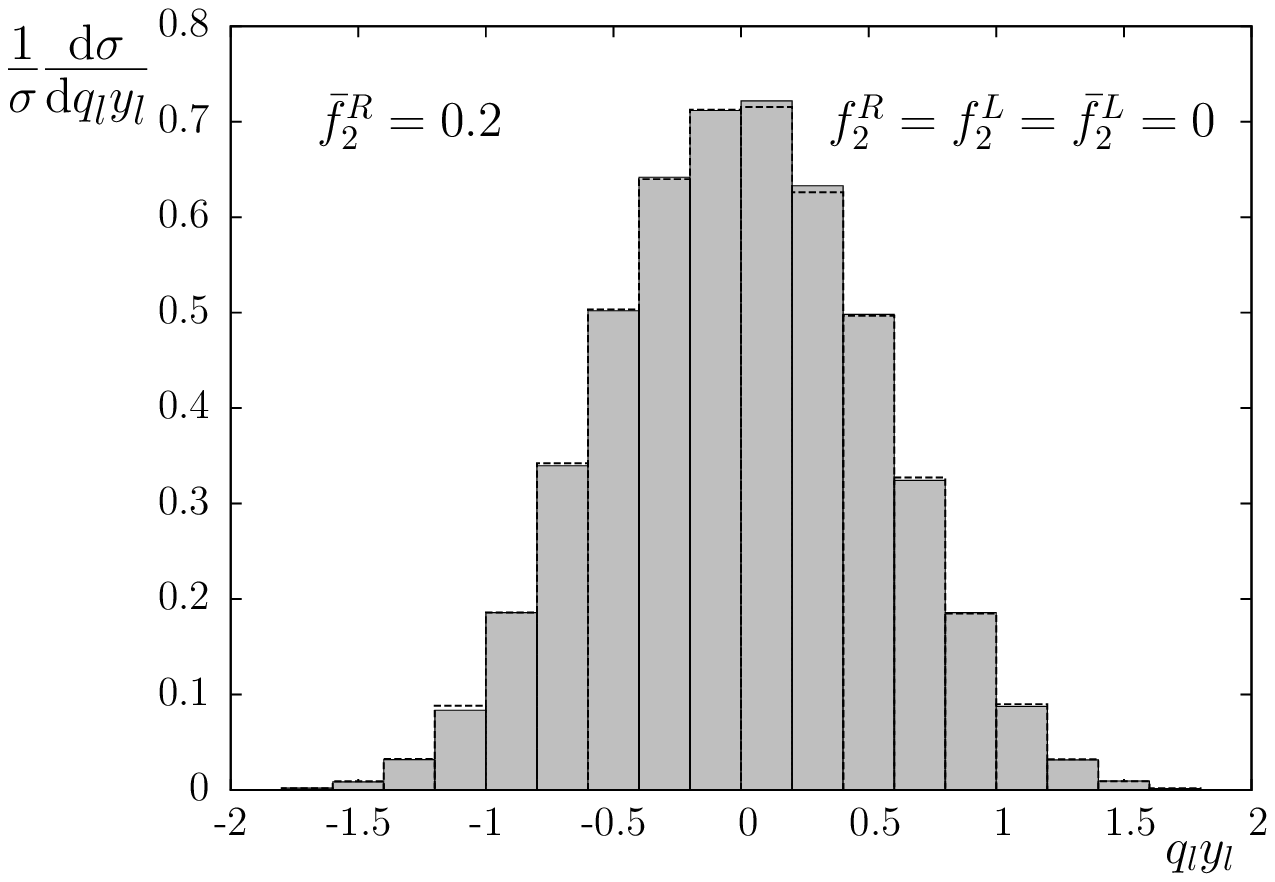}
\end{picture}
\end{center}
\vspace*{1.cm}
\caption{The charge-signed muon rapidity distribution for
different CP-even (upper raw) and CP-odd (upper raw) combinations 
of the tensor form factors of (\ref{lagr}). 
The vector form factors are fixed at their SM values of 
Eq.~(\ref{sm}). The SM result is shown in grey in each panel.
}
\label{figrapl}
\end{figure}

The charge-signed lepton rapidity distributions 
$1/\sigma\;{\rm d}\sigma/{\rm d}(q_ly_l)$, where $y_l$ is the rapidity of the 
charged lepton and $q_l$ is a sign of its electric charge, 
are plotted in Fig.~\ref{figrapl} for
different $CP$-even (upper raw) and $CP$-odd (upper raw) combinations 
of the tensor form factors of (\ref{lagr}).
The vector form factors are fixed at their SM values of 
Eq.~(\ref{sm}). The complete set of the lowest 
order Feynman diagrams is included in the calculation.
The SM result is shown in grey and the results
obtained with non zero tensor form factors are depicted with
boxes bounded by dashed lines in each panel.
In spite of the fact that the anomalous $Wtb$ coupling
changes the total cross section of the top quark pair production
by substantially altering the top quark width, the change 
in the charge-signed lepton rapidity distributions is hardly visible
in the plots.

\section{Summary}

The $t\bar t$ invariant mass dependent 
FBA of top quark production 
and the charge-signed rapidity distribution of the lepton originating
from the $W$ boson from top quark decay
at the Tevatron have been calculated to lowest order
taking into account the anomalous $Wtb$ coupling
of the most general form, with operators up to dimension five \cite{kane}.
It has been illustrated that even large values 
of the tensor form factors, exceeding the current limits \cite{afbD0}, 
have rather little influence on the FBA. 
Also the charge-signed rapidity distribution of the 
lepton is very little affected by different $CP$-even and $CP$-odd
combinations of the tensor
form factors within the current limits.

Acknowledgements: This work was supported in part by the Research Executive 
Agency (REA) of the European Union under the Grant Agreement number 
PITN-GA-2010-264564 (LHCPhenoNet).

\end{document}